\documentclass[%
 reprint,
 superscriptaddress,
 bibnotes,
 amsmath,amssymb,
 aps,
 prb,
 floatfix
]{revtex4-2}

\usepackage{graphicx}
\usepackage{dcolumn}
\usepackage{bm}
\usepackage{hyperref}
\usepackage{float}
\hypersetup{
  colorlinks   = true,
  urlcolor     = black,
  linkcolor    = black,
  citecolor   = blue
}
\usepackage{etoolbox}
\apptocmd{\sloppy}{\hbadness 10000\relax}{}{}

\begin{document}

\title{Unique Charge-density-wave Order and Fluctuation-induced Phonon Softening in the Optimally-doped Trilayer Cuprate Bi$_2$Sr$_2$Ca$_2$Cu$_3$O$_{10+\delta}$}

\author{S. Nakata}
\thanks{These authors contributed equally to this work.}
\affiliation{Department of Material Science, Graduate School of Science, University of Hyogo, Ako, Hyogo 678-1297, Japan}

\author{K. Yamamoto}
\thanks{These authors contributed equally to this work.}
\affiliation{NanoTerasu Center, National Institutes for Quantum Science and Technology (QST), Sendai, 980-8579, Japan}

\author{J. Okamoto}
\affiliation{National Synchrotron Radiation Research Center, Hsinchu 30076, Taiwan}

\author{D. Shiga}
\affiliation{Institute of Multidisciplinary Research for Advanced Materials (IMRAM), Tohoku University, Sendai 980-8577, Japan}

\author{R. Takahashi}
\affiliation{Department of Material Science, Graduate School of Science, University of Hyogo, Ako, Hyogo 678-1297, Japan}

\author{H. Y. Huang}
\affiliation{National Synchrotron Radiation Research Center, Hsinchu 30076, Taiwan}

\author{H. Kumigashira}
\affiliation{Institute of Multidisciplinary Research for Advanced Materials (IMRAM), Tohoku University, Sendai 980-8577, Japan}

\author{H. Wadati}
\affiliation{Department of Material Science, Graduate School of Science, University of Hyogo, Ako, Hyogo 678-1297, Japan}

\author{A. Fujimori}
\affiliation{National Synchrotron Radiation Research Center, Hsinchu 30076, Taiwan}
\affiliation{Department of Physics, University of Tokyo, Bunkyo-ku, Tokyo 113-0033, Japan}

\author{S. Ishida}
\affiliation{National Institute of Advanced Industrial Science and Technology, Tsukuba, Ibaraki 305-8568, Japan}

\author{H. Eisaki}
\affiliation{National Institute of Advanced Industrial Science and Technology, Tsukuba, Ibaraki 305-8568, Japan}

\author{D. J. Huang}
\affiliation{National Synchrotron Radiation Research Center, Hsinchu 30076, Taiwan}

\author{J. Miyawaki}
\affiliation{NanoTerasu Center, National Institutes for Quantum Science and Technology (QST), Sendai, 980-8579, Japan}

\author{H. Suzuki}
\email[]{Contact author: hakuto.suzuki@tohoku.ac.jp}
\affiliation{Frontier Research Institute for Interdisciplinary Sciences, Tohoku University, Sendai 980-8578, Japan}
\affiliation{Institute of Multidisciplinary Research for Advanced Materials (IMRAM), Tohoku University, Sendai 980-8577, Japan}
\date{\today}

\begin{abstract}
Charge-density-wave (CDW) order is a ubiquitous feature of cuprate superconductors that competes with superconductivity. However, its behavior remains poorly understood in trilayer compounds, which contain the optimal number of CuO$_2$ layers for superconductivity. Here, we report Cu $L_3$- and O $K$-edge resonant inelastic x-ray scattering (RIXS) measurements of optimally-doped Bi$_{2}$Sr$_{2}$Ca$_{2}$Cu$_{3}$O$_{10+\delta}$ (Bi2223), the trilayer cuprate with the highest superconducting transition temperature ($T_c$) among Bi-based cuprates. We observe CDW order with a single in-plane wavevector of ${\bm q}_{\text{CDW}} = (0.25, 0)$, demonstrating that a coherent CDW order is formed across the trilayers. We also resolve CDW fluctuations emanating from the static order, whose profiles differ between the Cu $L_3$ and O $K$ edges. The phonon softening occurs where these CDW fluctuations interact with the phonon branch, indicating that the RIXS phonon anomaly is driven by the coupling between CDW fluctuations and the phonon branch.
\end{abstract}

\maketitle

Cuprate superconductors continue to hold the highest superconducting transition temperature ($T_c$) among transition metal compounds, even after the discovery of superconductivity in iron pnictides and nickelates. A central focus in deciphering high-temperature superconductivity is the doping evolution of symmetry-broken orders, such as antiferromagnetic and charge-density-wave (CDW) states, and their competition with superconductivity \cite{Keimer.B_etal.Nature2015,Comin.R_etal.Annu.-Rev.-Condens.-Matter-Phys.2016}. For instance, the renowned suppression of $T_c$ in 1/8-doped La$_{2-x}$Ba$_x$CuO$_{4}$ is now understood as a consequence of the competition between superconductivity and charge-stripe order \cite{Tranquada.J_etal.Nature1995}, a phenomenon broadly realized in doped Mott insulators \cite{Imada.M_etal.Rev.-Mod.-Phys.1998}. However, the subsequent discovery of CDW phases across a wider class of underdoped cuprates \cite{Kohsaka.Y_etal.Science2007,Ghiringhelli.G_etal.Science2012} defies a unified description of the mechanism of CDW formation. Moreover, the recent identification of CDW order in the overdoped regime \cite{Peng.Y_etal.Nat.-Mater.2018} suggests that low carrier concentration is not a key requirement for the charge instability.

Despite the increasing diversity of CDW phenomena in cuprates, its coexistence with superconductivity is a universal phenomenon. The phase competition is evidenced by the reduction of the CDW peak intensity upon entering the superconducting state \cite{Blanco-Canosa.S_etal.Phys.-Rev.-Lett.2013}, as well as by the emergence of three-dimensional CDW order when superconductivity is suppressed by an external magnetic field \cite{Gerber.S_etal.Science2015,Chang.J_etal.Nat.-Commun.2016}. A natural question is whether static CDW order can survive in a system optimized for superconductivity. Within a homologous series of cuprates, the maximal $T_c$ is realized in compounds containing three CuO$_2$ sheets \cite{Iyo.A_etal.J.-Phys.-Soc.-Jpn.2007}. Thus, trilayer cuprates serve as an ideal platform to test the stability of CDW order against robust superconductivity. However, the detailed nature of the CDW in optimally doped (OP) Bi2223, which possesses the highest $T_c$ of 110 K \cite{Fujii.T_etal.J.-Cryst.-Growth2001}, remains unexplored.

CDW order also affects the phonon dispersion via electron-phonon coupling. Because the positively charged crystal lattice undergoes a concomitant deformation in response to the periodic modulation of the electron density, phonon softening is expected near the CDW wavevector \cite{Zhu.X_etal.Adv.-Phys.:-X2017}. While inelastic neutron scattering \cite{Reznik.D_etal.Nature2006} and non-resonant inelastic x-ray scattering \cite{Le-Tacon.M_etal.Nat.-Phys.2014,Graf.J_etal.Phys.-Rev.-Lett.2008} can also detect phonon softening, resonant inelastic x-ray scattering (RIXS) offers distinct advantages. Because the RIXS cross section of phonon excitations directly reflects the strength of electron-phonon coupling \cite{Devereaux.T_etal.Phys.-Rev.-X2016}, RIXS can selectively probe strongly coupled phonon modes. In addition, RIXS also reveals the CDW fluctuations emanating from the static CDW wavevector, which cause the ``RIXS phonon", whose softening can be distinct from the conventional effect of phonon self-energy \cite{Chaix.L_etal.Nat.-Phys.2017,Lee.W_etal.Nat.-Phys.2021,Huang.H_etal.Phys.-Rev.-X2021,Lu.H_etal.Phys.-Rev.-B2022}.

\begin{figure}[tbp]
  \centering
  \includegraphics[angle = 0, width = 0.48\textwidth, clip=true]{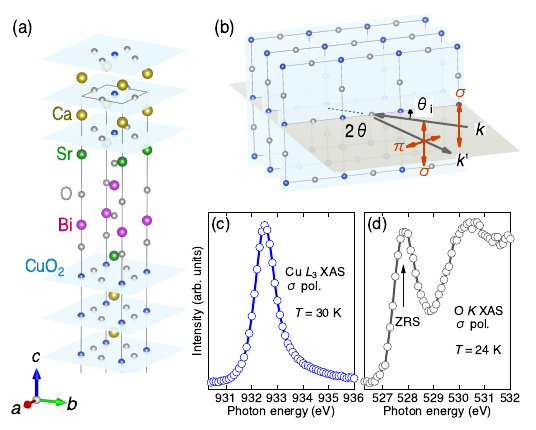}
  \caption{(a) Crystal structure of Bi$_2$Sr$_2$Ca$_2$Cu$_3$O$_{10+\delta}$ (Bi2223) \cite{Momma.K_etal.J.-Appl.-Cryst.2011}. (b) Scattering geometry of the resonant inelastic x-ray scattering (RIXS) experiment. The scattering angle was fixed at $2\theta$ = 150$^\circ$. The incident x-ray photons were $\sigma$-polarized, and the polarizations of the scattered photons were not analyzed.  (c) Cu $L_3$-edge x-ray absorption spectrum (XAS) collected with $\sigma$-polarized photons.  (d) O $K$-edge XAS collected with $\sigma$-polarized photons. The arrow indicates the absorption peak (527.9 eV) associated with the Zhang-Rice singlet (ZRS) states.} 
  \label{fig:1}
\end{figure}

In this Letter, we report CDW order, fluctuations, and RIXS phonon anomalies in the optimally doped trilayer cuprate Bi2223 with a high $T_c$ of 110 K \cite{SM}, the highest among the cuprates in which CDW order has so far been experimentally identified. Our Cu $L_3$- and O $K$-edge RIXS measurements consistently reveal short-range static CDW order at an in-plane wavevector of ${\bm q}_{\text{CDW}} = (0.25, 0)$ reciprocal lattice units (r.l.u.), and the CDW fluctuations emanating from the static order. The observation of this CDW deep within the superconducting state demonstrates that charge order can coexist with robust superconductivity even in a trilayer cuprate optimized for superconductivity. Moreover, the wavevector is nearly identical to those in optimally doped single- and double-layer Bi-based cuprates, despite the distinct electronic structures and layer-dependent doping in Bi2223. Notably, only a single CDW wavevector is observed despite the three inequivalent CuO$_2$ layers, indicating that the charge order is unique rather than being split into multiple, layer-dependent orders. Furthermore, the combination of the high $T_c$ and the high energy resolution of our RIXS setup allows us to clearly resolve the CDW fluctuations emanating from the static order. 

Figure \ref{fig:1}(a) shows the crystal structure of Bi2223, which contains three CuO$_2$ layers that are spatially separated by the Bi/Sr spacer layers. The Fermi surfaces, revealed by angle-resolved photoemission measurements \cite{Ideta.S_etal.Phys.-Rev.-Lett.2010,Kunisada.S_etal.Phys.-Rev.-Lett.2017,Luo.X_etal.Nat.-Phys.2023}, indicate that the two outer CuO$_2$ layers are overdoped, while the inner CuO$_2$ layer is underdoped. This doping imbalance arises because the hole carriers introduced from the spacer layers primarily enter the adjacent outer layers. Unlike YBa$_2$Cu$_3$O$_{6+x}$ (YBCO) \cite{Ghiringhelli.G_etal.Science2012} and the Hg-based cuprates \cite{Tabis.W_etal.Nat.-Commun.2014,Yu.B_etal.Phys.-Rev.-X2020}, charge order in single- and double-layer Bi-based cuprates prevails over a wide doping range, including the overdoped side \cite{Comin.R_etal.Science2014,Silva-Neto.E_etal.Science2014,Peng.Y_etal.Nat.-Mater.2018,Lu.H_etal.Phys.-Rev.-B2022}. It is therefore highly nontrivial whether the triple CuO$_2$ sheets in Bi2223 with distinct hole concentrations host a unique or multiple CDW orders. Moreover, Bi2223 possesses the highest $T_c=110$ K not only in the Bi-based family but also among all cuprates in which CDW order has been experimentally identified, with large superconducting gaps of different magnitudes in the inner and outer planes \cite{Ideta.S_etal.Phys.-Rev.-Lett.2010,Kunisada.S_etal.Phys.-Rev.-Lett.2017,Suzuki.H_etal.npj-Quantum-Mater.2018,Ideta.S_etal.Phys.-Rev.-Lett.2021,Luo.X_etal.Nat.-Phys.2023,Ideta.S_etal.Nat.-Commun.2025}. Bi2223 is therefore an ideal platform to study the stability of CDW order against superconductivity. Also, the large superconducting gap reduces the low-energy particle-hole excitations, which in turn suppresses the dissipation of the CDW fluctuations \cite{Lee.W_etal.Nat.-Phys.2021,Lu.H_etal.Phys.-Rev.-B2022}. This suppression of dissipation, together with the high energy resolution of our RIXS setup, allows the clear detection of CDW fluctuations in Bi2223, which can be difficult in low-$T_c$ cuprates with small superconducting gaps. 

To reveal the CDW-related phenomena in OP Bi2223, we employed Cu $L_3$- and O $K$-edge RIXS. Figure \ref{fig:1}(b) shows the scattering geometry of the RIXS experiment. We treat the electronic structure of Bi2223 as quasi-two-dimensional, and the in-plane momentum transfer ${\bm q}$ was scanned by changing the sample angle $\theta_i$, while the detector angle $2\theta$ was fixed at 150$^\circ$ to maximize the momentum transfer. Incident x-ray photons were $\sigma$-polarized, and both $\sigma$- and $\pi$-polarized scattered photons were collected with a detector. The Cu $L_3$- and O $K$-edge RIXS experiments were performed using the 2D-RIXS spectrometer at NanoTerasu \cite{Miyawaki.J_etal.J.-Phys.:-Conf.-Ser.2022,Yamamoto.K_etal.J.-Phys.:-Conf.-Ser.2025,Miyawaki.J_etal.Synchrotron-Radiat.-News2025,Yamamoto.K_etal.J.-Synchrotron-Rad.2026} and the AGS-AGM spectrometer at beamline 41A of Taiwan Photon Source \cite{Singh.A_etal.J.-Synchrotron-Rad.2021}, respectively. The total energy resolution was set to 32 meV at the Cu $L_3$ edge and 22 meV at the O $K$ edge. The instrument setup for the O $K$-edge RIXS was identical to that in Ref. \cite{Nakata.S_etal.Phys.-Rev.-B2025_plasmon}. In the following, we use the pseudotetragonal notation with lattice constants $a=b=3.85$~\AA\ and $c=37.16$ \AA\ \cite{Fujii.T_etal.J.-Cryst.-Growth2001}. The momentum transfer ${\bm q}=(H, K, L)$ is expressed in reciprocal lattice units $(2\pi/a, 2\pi/b, 2\pi/c)$. The details of the sample preparation and spectral fitting are described in the Supplemental Material \cite{SM}.

Figures \ref{fig:1}(c) and (d) present the x-ray absorption spectra (XAS) around the Cu $L_3$ and O $K$ edges, respectively, collected with $\sigma$-polarized x-rays. For the following RIXS measurements, we tuned the incident x-ray energy to the main absorption peak for the Cu $L_3$ edge, and to the transition peak associated with the Zhang-Rice singlet states \cite{Zhang.F_etal.Phys.-Rev.-B1988} at 527.9 eV (arrow) for the O $K$ edge. The latter condition enhances the charge response from the itinerant O $2p$ holes in hole-doped cuprates \cite{Nag.A_etal.Phys.-Rev.-Lett.2020,Singh.A_etal.Phys.-Rev.-B2022}.

\begin{figure*}[hbpt]
  \centering
  \includegraphics[angle = 0, width = 1\textwidth, clip=true]{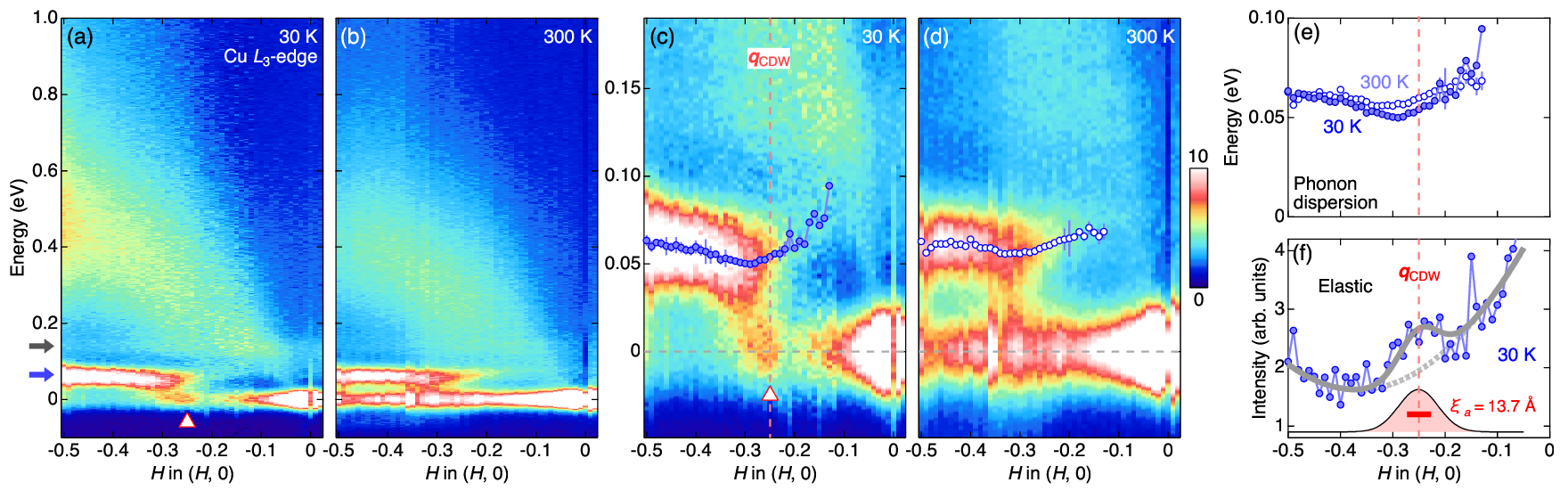}
  \caption{(a), (b) Cu $L_3$-edge RIXS intensity maps along the ${\bm q} = (H, 0)$ direction at $T = 30$ K and 300 K, respectively. The blue and gray arrows indicate the phonon and two-phonon excitations, respectively. The triangle indicates the static CDW wavevector ${\bm q}_{\text{CDW}}$ = (-0.25, 0).
  (c), (d) Expanded plots of the low-energy region highlighting the CDW order and fluctuations, and RIXS phonon branch. The circles indicate the fitted phonon peak energies. (e) Extracted phonon dispersions at 30 K and 300 K. (f) Momentum dependence of the quasi-elastic intensity at 30 K. The intensity profile is fitted to a Gaussian on top of a smooth background. The red horizontal bar indicates the inverse of the correlation length $\xi_a=13.7$ \AA\ of the CDW peak.} \label{fig:2}
\end{figure*}

We first present the Cu $L_3$-edge RIXS data of optimally doped Bi2223 along the in-plane momentum transfer $\bm{q} = (H, 0)$. Figures \ref{fig:2}(a) and (b) show RIXS intensity maps at $T = 30$~K and $300$~K, respectively. At 30 K, the elastic line shows a pronounced enhancement at $\bm{q}_{\text{CDW}} = (-0.25, 0)$ due to the formation of static CDW order (triangle). This wavevector agrees well with that in OP double-layer Bi$_2$Sr$_2$CaCu$_2$O$_{8+\delta}$ (Bi2212) \cite{Lee.W_etal.Nat.-Phys.2021,Lu.H_etal.Phys.-Rev.-B2022}. It is noteworthy that despite the distinct doping levels of the inner and outer CuO$_2$ layers in Bi2223, only a single CDW wavevector is observed. The RIXS spectra also contain a phonon excitation (indicated by the blue arrow) and a two-phonon excitation at higher energies (gray arrow). In analogy with Bi2212, the phonon branch is ascribed to the Cu-O bond-stretching phonons \cite{Graf.J_etal.Phys.-Rev.-Lett.2008}. While the phonon branch shows softening around the CDW wavevector as we detail below, the two-phonon dispersion is almost flat in ${\bm q}$ space, analogous to that in underdoped Bi2212 \cite{Scott.K_etal.Phys.-Rev.-B2024}. The broad dispersive feature below 1 eV is attributed to a combination of paramagnon excitations in the $\sigma$-$\pi$ scattering channel and the particle-hole charge continuum in the $\sigma$-$\sigma$ channel \cite{Minola.M_etal.Phys.-Rev.-Lett.2015}. At 30 K, the continuum excitations are pushed above the two-phonon energy except near the Brillouin zone center, whereas at 300 K they broaden and overlap with the phonon and two-phonon excitations. Below $T_c$, the $d$-wave superconducting gap ($\Delta$) opens on the Fermi surfaces, with a larger gap on the inner plane ($\Delta\sim 60$ meV) than on the outer planes ($\Delta\sim 40$ meV) \cite{Ideta.S_etal.Phys.-Rev.-Lett.2010,Kunisada.S_etal.Phys.-Rev.-Lett.2017,Ideta.S_etal.Phys.-Rev.-Lett.2021}. The superconducting gaps also suppress the low-energy particle-hole excitations and the corresponding RIXS spectral weight within the energy scale of $\sim 2\Delta$ \cite{Marra.P_etal.Phys.-Rev.-Lett.2013,Suzuki.H_etal.npj-Quantum-Mater.2018}. The suppression of the electron-hole continuum in the superconducting state reduces the dissipation of the CDW fluctuations \cite{Lee.W_etal.Nat.-Phys.2021,Lu.H_etal.Phys.-Rev.-B2022}, which allows the clear observation of the CDW fluctuations in Bi2223. 

To further explore the interaction between the phonon branch and the CDW fluctuations, we show expanded plots of the low-energy RIXS spectra in Figs. \ref{fig:2}(c) and (d). At 30 K, CDW fluctuations emanate from the static order at ${\bm q}_{\mathrm{CDW}}$ and show dispersive behavior in the larger $|H|$ region. The fluctuations intersect with the phonon branch, leading to softening of the phonon dispersion and enhancement of the phonon intensity in the region of $|H|\gtrsim 0.3$. It is noteworthy that the phonon-energy minimum is observed at $|H|=0.29$. This wavevector does not coincide with ${\bm q}_{\mathrm{CDW}}$ but approximately corresponds to the intersection of the CDW excitations and the phonon branch. This observation reinforces the notion that the softening of the phonon branch observed by RIXS is driven by the interaction between the phonon branch and quantum fluctuations of the CDW order via electron-phonon coupling \cite{Lee.W_etal.Nat.-Phys.2021,Huang.H_etal.Phys.-Rev.-X2021}. At 300 K, the CDW order is no longer clearly identifiable due to the increased elastic signal from, e.g., thermal diffuse scattering. Nevertheless, the flattening of the phonon dispersion suggests the suppression of the CDW order and fluctuations. 

The comparison of the phonon dispersions at 30 K and 300 K shown in Fig. \ref{fig:2}(e) reveals the relative weakness of the phonon anomaly in Bi2223. While the phonon dispersion at 300~K is relatively flat, the dispersion at 30~K shows only a small renormalization of $\Delta\omega \sim 7$~meV at $H=-0.29$. This softening, defined as the energy difference between the low- and room-temperature data, is much smaller than the phonon softening of $\sim 22$ meV observed in optimally doped Bi2212 \cite{Lee.W_etal.Nat.-Phys.2021,Lu.H_etal.Phys.-Rev.-B2022}.

To rigorously identify the correlation length of the CDW order, Fig. \ref{fig:2}(f) shows the fitted quasi-elastic intensity at 30 K. It exhibits a clear peak at $H=-0.25$. By fitting the momentum profile of the quasi-elastic intensity with a Gaussian function superimposed on a smooth background, we obtain a correlation length $\xi_a$ of 13.7 \AA\ for the CDW order. This value is shorter than the $4a$ periodicity of the order, and comparable to those reported in optimally doped Bi2212 \cite{Lee.W_etal.Nat.-Phys.2021,Lu.H_etal.Phys.-Rev.-B2022}, indicating that the CDW order in Bi2223 is similarly short-ranged.

\begin{figure}[hpt]
  \centering
  \includegraphics[angle = 0, width = 0.48\textwidth, clip=true]{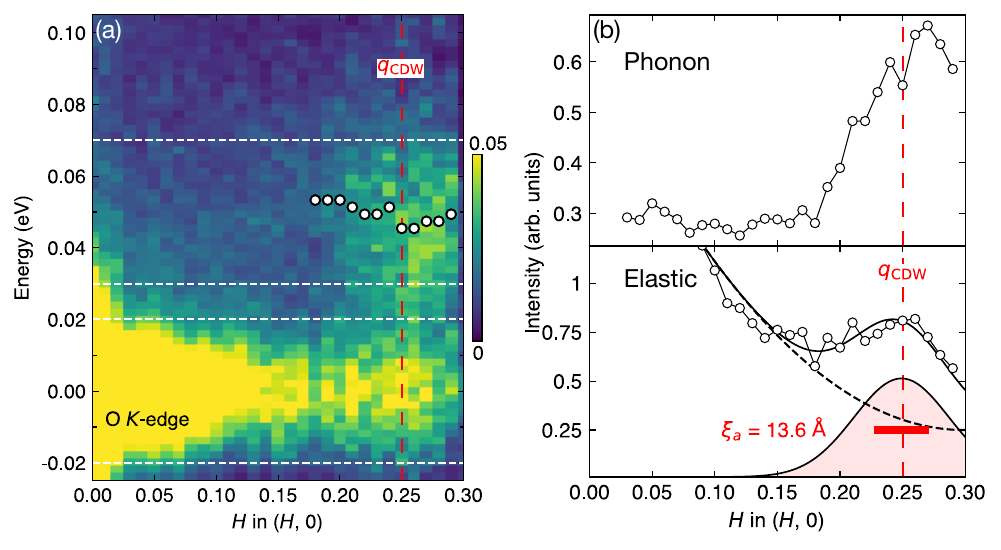}
  \caption{(a) Low-energy RIXS intensity map taken with O $K$-edge photons ($h\nu=527.9$ eV). The red dashed line indicates the static charge-density-wave vector ${\bm q}_{\text{CDW}}=(0.25, 0)$. The white circles represent the peak positions of phonon excitations. The error bars represent the standard deviation of the phonon energy. White dashed lines indicate the integration windows for the phonon intensity ($|\omega - 50\ \text{meV}|<20$ meV) and the elastic CDW intensity ($|\omega |<20$ meV). (b) Momentum dependence of the phonon intensity (top) and the elastic peak intensity (bottom) around ${\bm q}_{\text{CDW}}$. The elastic intensity profile is fitted to a Gaussian on top of a smooth background. The red horizontal bar indicates the inverse of the correlation length $\xi_a=13.6$ \AA\ of the CDW peak.} \label{fig:3}
\end{figure}

We next turn to the O $K$-edge RIXS data to examine the CDW phenomena from the perspective of the O~$2p$ electrons. Figure~\ref{fig:3}(a) shows the low-energy RIXS intensity map measured at $T = 24$~K with O $K$-edge photons ($h\nu = 527.9$~eV), along the $\bm{q} = (H, 0)$ direction. A clear peak in the elastic line is again observed at
${\bm q}_{\mathrm{CDW}} = (0.25, 0)$ (red dashed line), consistent with the CDW wavevector
identified by the Cu $L_3$-edge measurements. The correlation length $\xi_a = 13.6$~\AA\ extracted from a Gaussian fit to the elastic intensity [Fig.~\ref{fig:3}(b), lower panel] is in excellent agreement with the value obtained at the Cu $L_3$ edge, confirming that the CDW modulation simultaneously involves both the Cu~$3d$ and O~$2p$ orbitals. This multiorbital nature of the static CDW order is consistent with the result for single-layer Bi$_2$Sr$_{2}$LaCuO$_{6+\delta}$ (Bi2201)~\cite{Li.J_etal.Proc.-Natl.-Acad.-Sci.-U.S.A.2020}. The observation of static CDW order down to 24 K, well below the superconducting transition $T_c$ of 110 K, points to the ubiquity of charge order in cuprate superconductors coexisting with robust superconductivity. 

Despite the equivalence of the static CDW order observed in the Cu $L_3$- and O $K$-edge data, we find a notable difference in the CDW fluctuations and RIXS phonon softening. The O $K$-edge data also show a phonon branch near $\omega \sim 0.05$~eV around ${\bm q}_{\mathrm{CDW}}$, whose peak positions are indicated by white circles in Fig.~\ref{fig:3}(a). The dispersion exhibits an energy minimum near ${\bm q}_{\mathrm{CDW}}$, in contrast to the minimum at $|H|=0.29$ in the Cu $L_3$-edge data. In addition, the momentum dependence of the phonon intensity is distinct from that observed in the Cu $L_3$-edge data. The upper panel of Fig.~\ref{fig:3}(b) shows the phonon intensity integrated over $|\omega - 50~\mathrm{meV}| < 20$~meV. It exhibits a pronounced enhancement near ${\bm q}_{\mathrm{CDW}}$, in contrast to the enhancement of the spectral weight at larger $|H|$ observed in the Cu $L_3$ data.
 
The different locations of the phonon dispersion minima appear to originate from the distinct intensity profiles of the CDW fluctuations probed with the two resonances. While the CDW fluctuations in the Cu $L_3$-edge spectra show a steep linear dispersion from ${\bm q}_{\mathrm{CDW}}$ to the larger $|H|$ region [see Fig. \ref{fig:2}(c)], the O $K$-edge data show fluctuations that emanate almost vertically from ${\bm q}_{\mathrm{CDW}}$. This contrast indicates that RIXS at the two edges probes the same CDW correlations but with different cross sections in the $({\bm q}, \omega)$ space, reflecting the different electron-phonon coupling strengths for the Cu $3d$ and O $2p$ orbitals \cite{Devereaux.T_etal.Phys.-Rev.-X2016}. 

This edge dependence has also been observed in La$_{2-x}$Sr$_{x}$CuO$_4$ (LSCO) \cite{Lin.J_etal.Phys.-Rev.-Lett.2020,Huang.H_etal.Phys.-Rev.-X2021,Tam.C_etal.Phys.-Rev.-B2026,Huang.H_etal.Nat.-Commun.2026}. At the Cu $L_3$ edge, the bond-stretching phonon softening is strongest at a momentum larger than the quasi-elastic CDW peak, while the phonon intensity follows the expected momentum dependence of the bond-stretching electron-phonon coupling without a distinct anomaly at the CDW wavevector \cite{Lin.J_etal.Phys.-Rev.-Lett.2020,Tam.C_etal.Phys.-Rev.-B2026}. In contrast, O $K$-edge RIXS resolves low-energy charge fluctuations that emanate vertically from the CDW wavevector and produces a phonon anomaly centered close to that wavevector \cite{Huang.H_etal.Phys.-Rev.-X2021,Huang.H_etal.Nat.-Commun.2026}. These results indicate that the difference in the cross sections of Cu $L_3$- and O $K$-edge RIXS leads to distinct momentum-space structures of the CDW fluctuations, which in turn produce different RIXS phonon anomalies. The present results for Bi2223 demonstrate that this edge dependence is a universal feature of cuprates.

Despite the differences in the CDW fluctuation dynamics, the unique CDW wavevector $\bm{q}_{\mathrm{CDW}} = (\pm 0.25, 0)$ observed by the two resonances provides insight into the origin of the charge instability. In double-layer Bi2212, the CDW wavevector is weakly doping dependent in the underdoped regime but saturates near a commensurate value of 0.25 r.l.u. ($4a$ periodicity) in the optimally doped and overdoped regions \cite{Lu.H_etal.Phys.-Rev.-B2022}. Meanwhile, the Fermi surface topology and the quasiparticle spectral weight in the antinodal regions change drastically with doping. The persistence of the identical 0.25 r.l.u. wavevector across different doping concentrations poses a challenge to the Fermi surface nesting as the key determinant of the CDW periodicity. This issue becomes even more pronounced in trilayer Bi2223, where the unit cell contains inequivalent CuO$_2$ planes with distinct carrier concentrations. If the CDW wavevector were strictly dictated by the carrier concentration of each Fermi surface and the nesting vectors near the antinodal regions, multiple wavevectors would be expected in trilayer compounds. The presence of a single wavevector ${\bm q}_{\mathrm{CDW}}=(0.25, 0)$ suggests that the interlayer coupling locks the charge modulation into a single coherent periodicity across the trilayer unit. Indeed, the wavevector and the shorter correlation length than the $4a$ periodicity is analogous to the behavior observed in the optimally doped and overdoped regions of Bi2212 \cite{Lu.H_etal.Phys.-Rev.-B2022}. If the underdoped inner CuO$_2$ layer with a hole concentration of $\sim$ 7 \% \cite{Ideta.S_etal.Phys.-Rev.-Lett.2010,Ideta.S_etal.Nat.-Commun.2025} were solely responsible for the unique CDW order in Bi2223, a longer correlation length would be expected. Instead, the observed short correlation length is in line with the average hole concentration of $\sim$ 18 \% in Bi2223 \cite{Ideta.S_etal.Phys.-Rev.-Lett.2010,Ideta.S_etal.Nat.-Commun.2025}. Also, the gapped behavior of the plasmon branches of Bi2223 near the Brillouin zone center \cite{Nakata.S_etal.Phys.-Rev.-B2025_plasmon} highlights the crucial role of intra-trilayer electron hopping. These observations suggest that the interlayer electron hopping gives rise to a coherent CDW order across the trilayers.


Finally, although we have clearly identified the CDW fluctuations and concomitant RIXS phonon softening in Bi2223, the magnitude of the softening in the Cu $L_3$ data is much smaller than that in underdoped Bi2201 \cite{Li.J_etal.Proc.-Natl.-Acad.-Sci.-U.S.A.2020} and Bi2212 near optimal doping \cite{Chaix.L_etal.Nat.-Phys.2017,Lee.W_etal.Nat.-Phys.2021,Lu.H_etal.Phys.-Rev.-B2022}. The strength of CDW fluctuations that cause phonon softening may not be correlated with the strength of the CDW order, but rather linked to the proximity to the quantum critical point (QCP). Indeed, recent RIXS measurements on LSCO revealed quantum-critical scaling of the inverse correlation length \cite{Huang.H_etal.Nat.-Commun.2026}. The weaker phonon softening in OP Bi2223 may indicate that its electronic state is located away from the QCP where the dynamical CDW fluctuations are maximized. Systematic doping-dependent measurements of the CDW order and fluctuations will clarify this issue.

In conclusion, our Cu $L_3$- and O $K$-edge RIXS measurements establish that OP Bi2223, which has the highest $T_c=110$ K among the Bi-based cuprates, hosts a unique short-range CDW order at ${\bm q}_{\text{CDW}} = (0.25, 0)$. Notably, a coherent CDW order is stabilized across trilayers with inequivalent CuO$_2$ layers with distinct hole concentrations. We also clearly resolve the CDW fluctuations emanating from the static order. The distinct CDW fluctuation profiles and the RIXS phonon minima observed at the two resonances indicate that the RIXS phonon profile is determined by the interaction between the phonon branch and CDW fluctuations.  


\textit{Acknowledgements.}---
We thank T. Tohyama and A. Damascelli for enlightening discussions.
This project was supported by Grants-in-Aid for Scientific Research from JSPS (KAKENHI) (Nos. JP22K13994, JP25K00014) and by the JST FOREST Program (Grant Number JPMJFR256Q). The Cu $L_3$-edge RIXS experiments were performed using the 2D-RIXS spectrometer at BL02U of NanoTerasu
with the approval of the Japan Synchrotron Radiation Research Institute (JASRI) (Proposal Nos. 2025A9006, 2025A9059, 2025B9023, 2025B9047, 2026A9018, 2026B9004, 2026B9009, 2026B9017). The O $K$-edge RIXS experiment was performed using the AGS-AGM spectrometer at 41A of Taiwan Photon Source.

\bibliography{Bi2223_CDW}

\end{document}


\title{Supplemental Material for\\
  Unique Charge-density-wave Order and Fluctuation-induced Phonon Softening in the Optimally-doped Trilayer Cuprate Bi$_2$Sr$_2$Ca$_2$Cu$_3$O$_{10+\delta}$}

  \author{S. Nakata}
  \affiliation{Department of Material Science, Graduate School of Science, University of Hyogo, Ako, Hyogo 678-1297, Japan}

  \author{K. Yamamoto}
\affiliation{NanoTerasu Center, National Institutes for Quantum Science and Technology (QST), Sendai, 980-8579, Japan}

  \author{J. Okamoto}
  \affiliation{National Synchrotron Radiation Research Center, Hsinchu 30076, Taiwan}

  \author{D. Shiga}
  \affiliation{Institute of Multidisciplinary Research for Advanced Materials (IMRAM), Tohoku University, Sendai 980-8577, Japan}

  \author{R. Takahashi}
  \affiliation{Department of Material Science, Graduate School of Science, University of Hyogo, Ako, Hyogo 678-1297, Japan}

  \author{H. Y. Huang}
  \affiliation{National Synchrotron Radiation Research Center, Hsinchu 30076, Taiwan}

  \author{H. Kumigashira}
  \affiliation{Institute of Multidisciplinary Research for Advanced Materials (IMRAM), Tohoku University, Sendai 980-8577, Japan}

  \author{H. Wadati}
  \affiliation{Department of Material Science, Graduate School of Science, University of Hyogo, Ako, Hyogo 678-1297, Japan}

\author{A. Fujimori}
\affiliation{National Synchrotron Radiation Research Center, Hsinchu 30076, Taiwan}
\affiliation{Department of Physics, University of Tokyo, Bunkyo-ku, Tokyo 113-0033, Japan}

  \author{S. Ishida}
  \affiliation{National Institute of Advanced Industrial Science and Technology, Tsukuba, Ibaraki 305-8568, Japan}

  \author{H. Eisaki}
  \affiliation{National Institute of Advanced Industrial Science and Technology, Tsukuba, Ibaraki 305-8568, Japan}

  \author{D. J. Huang}
  \affiliation{National Synchrotron Radiation Research Center, Hsinchu 30076, Taiwan}

  \author{J. Miyawaki}
  \affiliation{NanoTerasu Center, National Institutes for Quantum Science and Technology (QST), Sendai, 980-8579, Japan}

  \author{H. Suzuki}
  \affiliation{Frontier Research Institute for Interdisciplinary Sciences, Tohoku University, Sendai 980-8578, Japan}
  \affiliation{Institute of Multidisciplinary Research for Advanced Materials (IMRAM), Tohoku University, Sendai 980-8577, Japan}

  \maketitle

  \section{\NoCaseChange{Characterization of Bi$_{2}$Sr$_{2}$Ca$_{2}$Cu$_{3}$O$_{10+\delta}$ single crystals}}

  \begin{figure}[hbt]
    \centering
    \includegraphics[width=0.5\textwidth, clip=true]{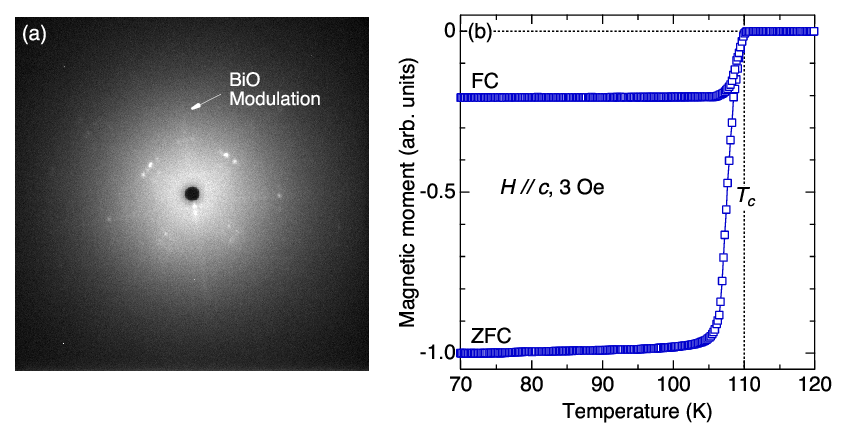}
    \caption{\label{fig:S1} (a) Laue diffraction pattern of the Bi2223 single
    crystal. (b) Temperature-dependent magnetic moment for the optimally doped
    Bi2223 sample ($T_{c}= 110$ K), measured under zero-field-cooled (ZFC) and
    field-cooled (FC) conditions with a 3 Oe magnetic field applied parallel to
    the $c$ axis.}
    \label{fig:S1}
  \end{figure}

  We determined the orientation of the optimally doped Bi$_{2}$Sr$_{2}$Ca$_{2}$Cu$_3$O$_{10+\delta}$
  (Bi2223) single crystal used for the RIXS measurements by Laue diffraction, as
  shown in Fig. \ref{fig:S1}(a). In this pattern, the linear feature arises from
  the charge modulation within the BiO planes along the ${\bm q}=(H, H, L)$ direction.

Figure \ref{fig:S1}(b) displays the temperature dependence of the magnetic moment. These measurements were performed under a 3 Oe magnetic field applied parallel to the $c$ axis using both zero-field-cooled (ZFC) and field-cooled (FC) protocols. The data are normalized to the value at 70 K in the ZFC process. The superconducting transition temperature ($T_{c}$) of 110 K was determined from the onset of the Meissner diamagnetic signal.
  \\

\section{\NoCaseChange{Fitting procedure of Cu $L_{3}$-edge RIXS spectra}}

\begin{figure*}[t]
    \centering
    \begin{minipage}[t]{0.49\textwidth}
    \centering
    \includegraphics[angle=0, width=\linewidth, clip=true]{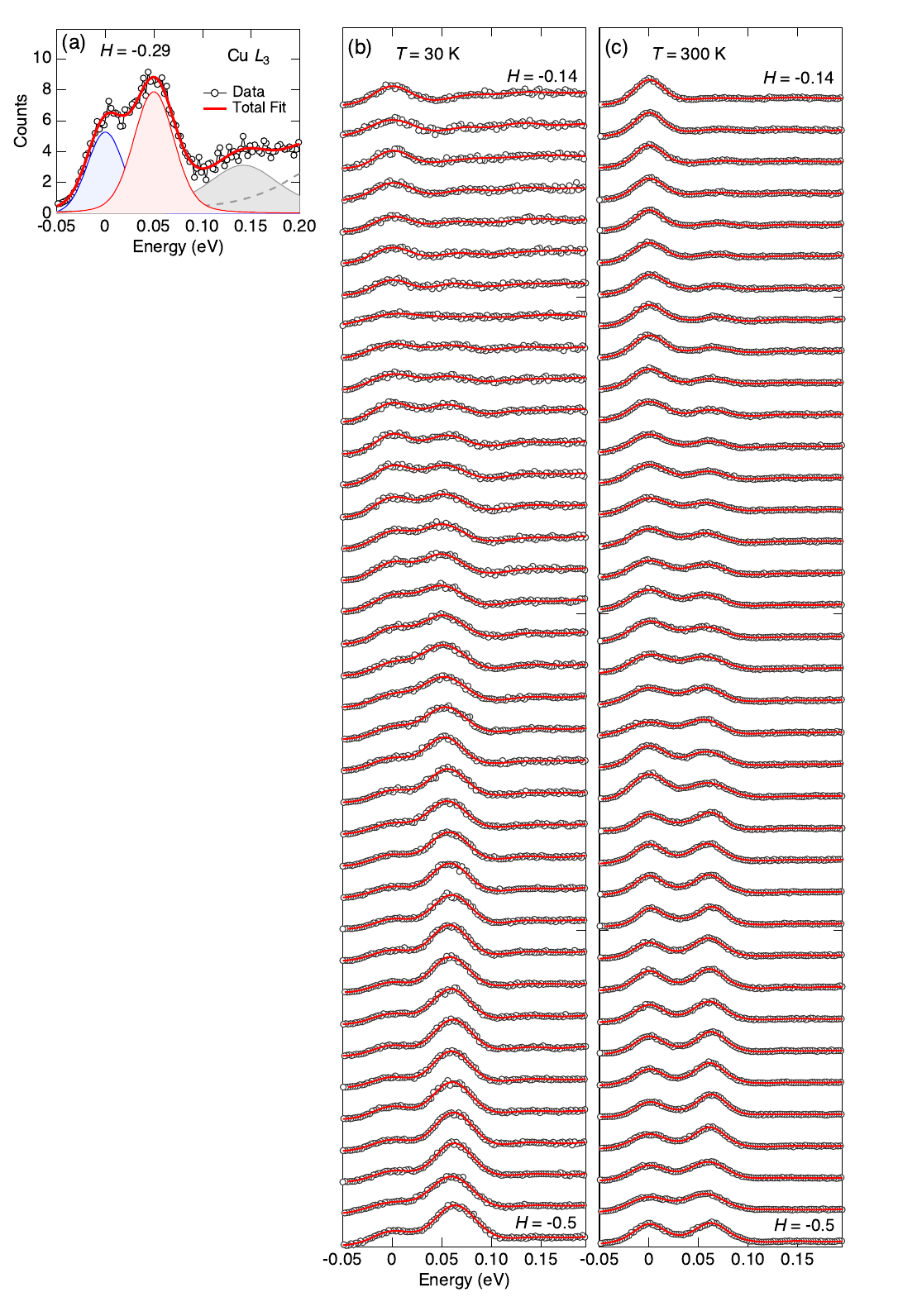}
    \caption{(a) A representative Cu $L_{3}$-edge RIXS spectrum of optimally doped
    Bi2223 at ${\bm q}= (-0.29, 0)$. The spectrum was fitted to three Voigt profiles
    corresponding to the elastic line (blue), phonon (red), and two-phonon excitation (gray),
    on top of a smooth background (dashed curve). (b),(c) Spectral fitting of the Cu $L_3$-edge
    RIXS spectra along the ${\bm q}= (H, 0)$ direction at $T=30$ and $300$ K,
    respectively.}
    \label{fig:S2}
    \end{minipage}
    \hfill
    \begin{minipage}[t]{0.41\textwidth}
    \centering
    \includegraphics[angle=0, width=\linewidth, clip=true]{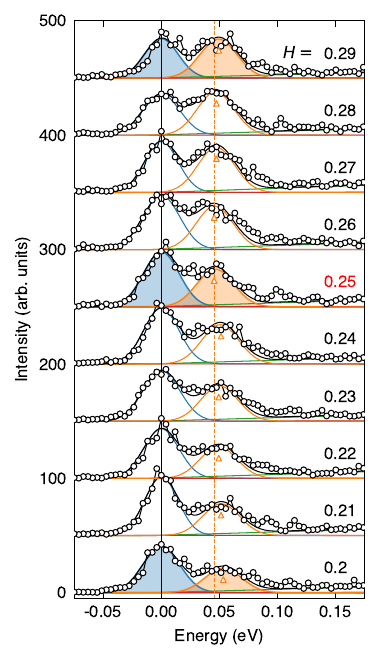}
    \caption{Spectral fitting of O $K$-edge RIXS spectra of optimally doped Bi2223. The elastic lines
    (blue) and phonon peaks (orange) are fitted to Voigt functions. The
    orange triangles indicate the determined phonon energy. The orange dashed vertical
    line indicates the phonon-energy minimum at the CDW wavevector ${\bm q}
    _{\text{CDW}}=(0.25, 0)$. }
    \label{fig:S3}
    \end{minipage}
  \end{figure*}

To quantitatively determine the phonon dispersion and quasi-elastic intensity, we fitted the low-energy Cu $L_{3}$-edge RIXS spectra. A representative spectrum measured at ${\bm q}=(-0.29, 0)$ is shown in Fig. \ref{fig:S2}(a).
The spectrum was fitted to three Voigt functions representing the elastic line (blue), the bond-stretching phonon (red), and the two-phonon
excitation (gray), on top of a smooth background that accounts for a weak continuum (dashed curve). The instrumental energy resolution of 32 meV was incorporated
as the Gaussian broadening of each Voigt function. 
Figures \ref{fig:S2}(b) and \ref{fig:S2}(c) show the fitting results for the series of Cu $L_{3}$-edge RIXS spectra measured along the ${\bm q}=(H,0)$ direction at $T=30$ and $300$ K, respectively. The extracted phonon peak positions were used as the RIXS phonon dispersions presented in Fig. 2 of the main text. The integrated intensity of the elastic component was employed to evaluate the CDW peak profile and the corresponding correlation length.

\section{\NoCaseChange{Fitting procedure of O $K$-edge RIXS spectra}}

The O $K$-edge spectra were analyzed in the same manner as the Cu $L_{3}$-edge data. The quasi-elastic line and phonon peak were fitted to Voigt profiles, with the
  instrumental resolution (22 meV) included as the Gaussian broadening. 
  The quasi-elastic intensity (blue) shows a maximum at $\bm q_{\text{CDW}}=(0.25,0)$, corroborating the presence of CDW order, as observed in the Cu $L_{3}$-edge measurements. The extracted phonon energies (orange triangles in Fig. \ref{fig:S3}) exhibit a minimum at the same CDW wavevector ${\bm q}_{\text{CDW}}$, as
  marked by the dashed vertical line.

  \bibliography{Bi2223_CDW}